\documentclass[%
  reprint,
superscriptaddress,
showpacs,preprintnumbers,
nofootinbib,
 amsmath,amssymb, 
 aps,
 prd,
 longbibliography
]{revtex4-1}

\usepackage{cancel}
\usepackage{accents}
\usepackage{mciteplus,slashed}
\usepackage{amssymb,cancel,amsmath}
\usepackage{dcolumn}
\usepackage{bm}
\usepackage[caption=false]{subfig}
\usepackage{appendix}
\usepackage{feynmp-auto}
\usepackage[T1]{fontenc}	
\usepackage{csvsimple}
\usepackage[colorlinks=true,citecolor=blue,linkcolor=blue, allcolors=blue]{hyperref}
\usepackage[section]{placeins}
\usepackage[capitalise]{cleveref}
\usepackage{booktabs}
\usepackage{graphicx}
\usepackage{mathrsfs}
\usepackage[utf8]{inputenc}
\usepackage{siunitx}
\usepackage{aas_macros}

\usepackage[dvipsnames]{xcolor}
\usepackage[normalem]{ulem}

\definecolor{darkgreen}{rgb}{0.0, 0.2, 0.13}
\definecolor{bostonuniversityred}{rgb}{0.8, 0.0, 0.0}

\begin{document}

\preprint{\hfill FERMILAB-PUB-20-261-AE-PPD-T}

\title{Is GW170817 a Multimessenger Neutron Star-Primordial Black Hole Merger?}

\author{Yu-Dai Tsai}
\email{ytsai@fnal.gov}
\affiliation{Cosmic Physics Center, Fermi National Accelerator Laboratory, P. O. Box 500, Batavia, IL 60510, USA}
\affiliation{Theory Department, Fermi National Accelerator Laboratory, P. O. Box 500, Batavia, IL 60510, USA}
\affiliation{Kavli Institute for Cosmological Physics, University of Chicago, Chicago, IL 60637, USA}
\author{Antonella Palmese}
\email{palmese@fnal.gov}
\affiliation{Cosmic Physics Center, Fermi National Accelerator Laboratory, P. O. Box 500, Batavia, IL 60510, USA}
\affiliation{Kavli Institute for Cosmological Physics, University of Chicago, Chicago, IL 60637, USA}
\author{Stefano~Profumo}
\email{profumo@ucsc.edu}
\affiliation{Department of Physics, University of California Santa Cruz, and 
Santa Cruz Institute for Particle Physics, 1156 High St., Santa Cruz, CA 95064, USA}
\author{Tesla Jeltema}
\email{tesla@ucsc.edu}
\affiliation{Department of Physics, University of California Santa Cruz, and 
Santa Cruz Institute for Particle Physics, 1156 High St., Santa Cruz, CA 95064, USA}

\date{\today}

\begin{abstract}

We investigate the possibility of the gravitational-wave event GW170817 being a light, solar-mass black hole (BH) - neutron star (NS) merger. We explore two exotic scenarios involving primordial black holes (PBH) that could produce such an event, taking into account available observational information on NGC 4993. First, we entertain the possibility of dynamical NS-PBH binary formation where a solar-mass PBH and a NS form a binary through gravitational interaction. We find that while dynamical NS-PBH formation could account for the GW170817 event, the rate is highly dependent on unknown density contrast factors and could potentially be affected by galaxy mergers. We also find that PBH-PBH binaries would likely have a larger merger rate, assuming the density contrast boost factor of an order similar to the NS-PBH case. These exotic merger formations could provide new channels to account for the volumetric rate of compact-object mergers reported by LIGO/Virgo. Secondly, we consider the case where one of the NS's in a binary NS system is imploded by a microscopic PBH. We find that the predicted rate for NS implosion into a BH is very small, at least for the specific environment of NGC 4993. We point out that similar existing (e.g. GW190425 and GW190814) and future observations will shed additional light on these scenarios.

\end{abstract}

\maketitle 

\section{Introduction}

The discovery of the gravitational wave (GW) event GW170817 \cite{TheLIGOScientific:2017qsa} and its electromagnetic (EM) counterpart \cite{GBM:2017lvd}, has ushered the era of multi-messenger astrophysics, enabling the use of both EM and GW observations to explore cosmology \cite{Schutz:1986gp,Abbott:2017xzu,DES:2019irc,Palmese:2019ehe,Palmese:2020kxn,Palmese:2020aof} and fundamental physics \cite{Bramante:2017ulk,Tsai:2018qoa,Authors:2019qbw}.
In particular, the fact that a binary compact-object merger can be located within its host galaxy (e.g. NGC 4993 for GW170817) allows for an in-depth investigation of its various properties given its environment, which can be combined with rich observations from various wavelengths. 

GW170817 is classified as a binary neutron star (BNS) merger; however, the nature of GW170817 and similar mergers (e.g. GW190425 \citep{190425,Kyutoku:2020xka,Han:2020qmn}) is still under investigation. The identification of the kilonova (e.g. \cite{Soares-Santos:2017lru}) associated with GW170817 suggests that there was at least one neutron star (NS) involved in the merger. The possibility of GW170817 being a neutron star--black hole (NSBH) merger is first considered in \cite{Hinderer:2018pei}, which finds a $\lesssim 40 \%$ chance for the event, and further studies show that NSBH can account for the multimessenger signatures involving AT2017gfo and GRB170817A \cite{Coughlin:2019kqf}. The uncertainty on the origin of GW170817 also extends to the possible formation channels. 
Its host galaxy, NGC 4993, is an old elliptical galaxy with no recent signs of star formation. \cite{Belczynski2} finds that assuming this type of old stellar population, standard models of BNS formation may not be able to explain the rate of mergers measured by LIGO/Virgo, unless GW170817 was a statistical coincidence. It is thus interesting to consider additional formation channels or alternative classes of compact-object components.

Dark matter (DM) continues to be one of the deepest mysteries in modern physics. Primordial black holes, or PBHs \cite{Carr:1974nx,Meszaros:1974,Carr:1975}, result from large overdensities in the early universe, rather than through standard stellar evolution, and they have been proposed as possible DM candidates (e.g. \cite{Clesse:2015wea}).
While it is hard for astrophysical BHs to form with one or two solar masses, a PBH can form with a NS or solar mass  \cite{Byrnes:2018clq,Carr:2019kxo,Davoudiasl:2019ugw}; the detection of sub-Chandrasekhar mass black hole mergers would unmistakably lead to the discovery of this class of exotic compact objects \cite{Shandera:2018xkn, Authors:2019qbw, Lehmann:2020bby}. PBHs can have various effects on astrophysical objects, including the formation of bound states with other PBHs \cite{Raidal:2017mfl}, or the implosion of compact objects into low-mass BHs \cite{Capela:2013yf,Fuller:2017uyd,Bramante:2017ulk} which otherwise would rarely exist through standard BH formation channels.

In this paper, we discuss three new aspects related to mergers of compact objects:
\begin{enumerate}
	\item GW170817 could be a NS-PBH merger.
	\item The first estimation of NS-PBH dynamical formation (based on the astrophysical environment of NGC 4993 as an illustration).
	\item An alternative scenario: a microscopic PBH has imploded a NS in the GW170817 merger event.
\end{enumerate}
\newpage

First, we discuss the dynamical formation of a binary composed by a neutron star (NS) and a solar-mass PBH in the host galaxy of GW170817 through two-body gravitational interactions. This is the first work to estimate the NS-PBH dynamical formation, to the best of our knowledge.
We suggest that if BNS formation is dynamically driven in NGC 4993, as discussed in \cite{palmese}, NS-PBH and PBH-PBH formation also would be.
This scenario could, therefore, provide an interesting alternative to address some of the questions posed in \cite{palmese,Belczynski}, where it was found that it is unlikely for a binary NS formed through standard stellar evolution to merge in an old, elliptical galaxy like NGC 4993. 
Secondly, we discuss the possibility of a binary NS system, with one NS imploded into a micro-sized PBH, to yield an event compatible with GW170817. 

The scenarios we consider in this work can be applied to other mergers comprising solar-mass compact objects \cite{190425,Abbott:2020khf}, but GW170817 is unique given the multi--messenger observations and knowledge of its environment that are currently unavailable for other mergers.

This paper is organized the follows. In Sec. \ref{Sec:NGC_4993}, we describe the environment of GW170817. In Sec. \ref{sec:dynamical}, we show our derivation of the rate for NS-PBH dynamical formation. The scenario involving a NS imploded by a PBH is described in Sec. \ref{sec:implosion}. In Sect. \ref{sec:prospects}, we briefly mention the possibility of GW190425 \cite{190425} and GW190814 \cite{Abbott:2020khf} also containing a PBH, and mention future prospects.
We assume $c=1$ unless otherwise stated.

\section{The host galaxy of GW170817:\\ NGC 4993}
\label{Sec:NGC_4993}

NGC 4993 is a nearby ($z = 0.009680 \pm 0.000150$) early–type galaxy with stellar mass $M_\star=(3.8\pm0.20)\times 10^{10}~M_\odot$, with $i$-band Sérsic index $n$ = 4.0 and low asymmetry. Despite the fact that these results are consistent with a passively evolving, red and dead galaxy, the presence of shells and dust lanes indicate that it recently underwent a galaxy merger, as we describe in more detail in Section \ref{sec:merger}. 

From the perspective of BNS formation, the association of GW170817 as a BNS merger with an old elliptical galaxy, like NGC 4993, is puzzling \cite{palmese,Belczynski,Belczynski2}. The galaxy does not show recent signs of star formation, and the median inferred stellar population age is 11 Gyr \cite{palmese}. If the binary NS that merged was formed through the isolated binary scenario, this would imply a surprisingly long delay time\footnote{The delay time, in this case, is defined as the time between the formation of the binary and its coalescence.} of $\sim 11$ Gyr.

Assuming a flat $\Lambda$CDM cosmology with $H_0=70~{\rm km~s^{-1}~Mpc^{-1}}$ and $\Omega_m=0.3$, one can estimate that the merger related to GW170817 happened at a projected distance of $\sim 2$ kpc from the center of NGC 4993, where there is an abundance of both pulsars and DM \cite{levan}.

\subsection{Neutron Star/Pulsar Distribution}

To estimate the NS population in NGC 4993, we use the spherically-symmetric distribution of progenitors from
\cite{P90}
\begin{equation}
\label{eq:nsdist}
    \rho_{\rm NS}(r)=\rho^0_{\rm NS}\frac{r}{R_0}\exp(-r/R_0), 
\end{equation}
with $R_0=4.5$ kpc, and the normalization $\rho^0_{\rm NS}$ set such that the total integrated number of NS is equal to $N_{\rm NS}=6.7\times 10^7$. We estimate the former total number of NSs by computing the NS formation rate per stellar mass at time $t$ for NGC 4993:
\begin{equation}
R_{NS}(t) = \int dM_\star \Phi(M_\star)\Psi(t_\star)\Theta_{NS}(M_\star)\, ,\label{eq:rns}
\end{equation}
where $\Phi(M_\star)$ is the initial mass function (here we adopt the one from \cite{Chabrier:2003}), $\Psi(t_\star)$ is the best fit star formation history from \cite{palmese}, $\Theta_{NS}(M_\star)$ is a top--hat function that selects stars that will become NSs, i.e. in the mass range $8\, M_\odot<M<20 \,M_\odot$. The time $t_\star$ when the progenitor of the NS was formed satisfies $t = t_\star+t_{\rm life}$, where $t_{\rm life}\sim0.02$ Gyr is the lifetime of the progenitor before becoming a NS. We then integrate Eq. (\ref{eq:rns}) over time and multiply it by the total stellar mass of NGC 4993, in order to recover the total number of NS formed in the galaxy. The fraction of NSs formed in binaries ($\sim 0.002$ \cite{vangioni}), and thus possibly merging to form a BH, is a negligible fraction of the number provided above.

We varied those parameters quite generously in calculating the impact on the merger rates we present below, since no reliable NS population information is available for NGC 4993. Our work could motivate a more comprehensive study of the NS distribution based on the specifics of NGC 4993, especially given its elliptical morphology and light profile.

\subsection{Dark Matter Distribution}
\label{sec:DM_dist}

The mass of the NGC 4993 DM halo is estimated to be $M_{\rm halo,\;NGC4993} = 194^{+120}_{-70} \times 10^{10} M_\odot$ \cite{Ebrova:2018gtz}. In order to compute the rates described in Section \ref{sec:merger}, we need to estimate the radial DM profile for this galaxy. We assume a Navarro, Frenk and White (NFW \cite{Navarro:1996gj}) density profile, and the concentration-mass relation of Ref.~\cite{Dutton:2014xda}, which implies a concentration $c_{200}=7.52^{+2.17}_{-2.07}$. This implies a scale radius $R_s=20.7$ kpc. The radial velocity dispersion distribution we employ is derived using velocity dispersion maps from \cite{levan,hjort}, that have been derived using Integral Field Spectroscopy from MUSE on the Very Large Telescope. 
We further extend the velocity dispersion profile out to larger radii than what is observed by MUSE, following the classic results of Ref.~\cite{Lokas:2000mu} for the velocity dispersion as a function of radius in a given NFW profile under the assumption of isotropic orbits (their Eq.~(14)).

\subsection{Binary Formation\\ \& Galaxy Merger}\label{sec:merger}

It was noted above that the standard field formation for the BNS could be disfavored because NGC 4993 only hosts old stellar populations \cite{palmese,Belczynski}.
The expected rate of BNS mergers from this formation scenario can be derived for this galaxy following \cite{palmese}. Using the observed star formation history of NGC 4993, and time delay distribution from Milky Way constraints and simulations, it was found that the rate is $\sim 6$ Myr$^{-1}$ for this galaxy. 

Similarly, one can utilize stellar population synthesis models together with host galaxies' information to infer a volumetric rate of events from elliptical galaxies that can be directly compared to the one measured by LIGO/Virgo (see \cite{190425} for a recent estimate), as done in \cite{Belczynski}. They find that the rate from stellar evolution models is $\sim 3$ orders of magnitude lower than the one inferred by LIGO/Virgo, while the merger rate from binaries formed through dynamical interactions in globular clusters or nuclear clusters is $\sim6$ orders of magnitude lower than the LIGO/Virgo rate. These suggest that either GW170817 is a statistical coincidence, or there exist another formation mechanisms for these binaries.

It was postulated that dynamical formation and galaxy mergers could help explain the puzzle of the BNS rate \cite{palmese}.
The morphological features indicate that NGC 4993 underwent a galaxy merger \cite{palmese,Ebrova:2018gtz}. 
It was discussed that a secondary, less massive galaxy was accreted by a more massive galaxy, and the apocenters of in-falling material from the secondary are seen today as shells. From the position of the shells, it is possible to constrain the time of the galactic merger to be of the order of $\sim 200-400$ Myr before the GW coalescence \cite{palmese,Ebrova:2018gtz}.
It was noted that GW170817 was located in one of the shells (at a projected distance of $10.6''$, or $\sim 2$ kpc, from the center of the galaxy), which is a relic of the galactic merger, indicating that the binary formation could be related to such merger.

Given that dynamical interactions allow for more recent BNS formation independent of the star formation rate, the delay time can be reduced compared to the field BNS formation scenario. One can postulate that the galaxy merger also plays a role in enhancing the dynamical binary formation rate. Future simulations could potentially answer this question \cite{precite}. 

\section{NS-PBH Dynamical Formation}\label{sec:dynamical}

In this section, we describe the first estimation of the dynamical formation of NS-PBH binaries. Dynamical formation is one of the most promising formation channels for a NS-(solar mass) PBH merger in NGC 4993. Through this formation channel, we also consider PBH-PBH binary formation for comparison and future projections, since these PBH-PBH binaries could be interesting for other GW events containing solar-mass compact objects. We only consider dynamical formation through two-body gravitational interaction, and leave a more complete study, involving N-body simulations, for future work.

We are interested in PBH in the NS or solar mass ranges (more specifically the mass of one of the compact objects in GW170817).
For GW170817, the masses of the component compact objects are determined in \cite{Monitor:2017mdv}. The primary component has mass $m_1$ between 1.36 and 2.26 solar mass, and the secondary has mass $m_2$ between 0.86 and 1.36 solar mass, both at 90\% credible interval (CI).
The limit of PBH abundance in these mass ranges is between 0.01 and 0.1 of the total DM abundance \cite{1993Natur.366..242H,Tisserand:2006zx,Dong:2007px,Oguri:2017ock,Kelly:2017fps,Zumalacarregui:2017qqd}; however, the constraints on this mass range depend on a variety of assumptions (also, see \cite{Raidal:2017mfl}). 

In order for the scenario proposed here to be an interesting alternative to standard BNS formation scenarios, the rate for NS-PBH mergers should be of the order of 6 Myr$^{-1}$, as this is the rate derived for field BNS binary mergers in NGC 4993 \cite{palmese}. To match the NS-PBH rate to the
volumetric rate inferred by LIGO/Virgo solar-mass compact-object mergers by summing up potential contributing galaxies (as discussed in Sec. \ref{sec:merger}), one may need an even higher NS-PBH rate. Our analysis can be easily scaled for that purpose.
Note that the galaxy merger could play a role in enhancing the dynamical formation of binaries, however, a model that takes into account the galaxy merger does not currently exist to our knowledge. We therefore only consider dynamical formation within the galaxy as if it were isolated.

We estimate the NS-PBH and PBH-PBH merger rates in the late universe using the formalism of Ref.~\cite{Raidal:2017mfl}, which we briefly summarize here: binaries are formed in two-body interactions via the loss of energy due to gravitational radiation. In the Newtonian approximation, the velocity-averaged cross-section for binary formation can be cast as \cite{Mouri:2002mc, Raidal:2017mfl}
\begin{equation}
    \langle \sigma v\rangle(m_1,m_2)=A G_N^2 M^2\left(\frac{\eta}{v_{\rm rel}^9}\right)^{2/7},
\end{equation}
where $v_{\rm rel}$ is the relative velocity between the merging bodies, 
$G_N$ is Newton's gravitational constant, $M=m_1+m_2$ is the sum of the components' masses $m_1$ and $m_2$, $\eta=m_1m_2/M^2$, and $A$ is a constant that depends on the velocity distribution with $A\simeq 12.9$ for the assumed Maxwellian distribution. Assuming a monochromatic mass distribution for the merging bodies, and assuming spherical symmetry, the merger rate is calculated as
\begin{equation}
    R(m_1,m_2)=4\pi\int_0^{R_{\rm vir}} r^2dr \langle \sigma v\rangle(r,m_1,m_2)\frac{\rho_1(r)}{m_1}\frac{\rho_2(r)}{m_2},
\end{equation}
where $\rho_i$ is the mass  density distribution for species $i$, and $R_{\rm vir}$ is the virial radius. 

\begin{figure}
    \centering
    \includegraphics[width=9.0 cm]{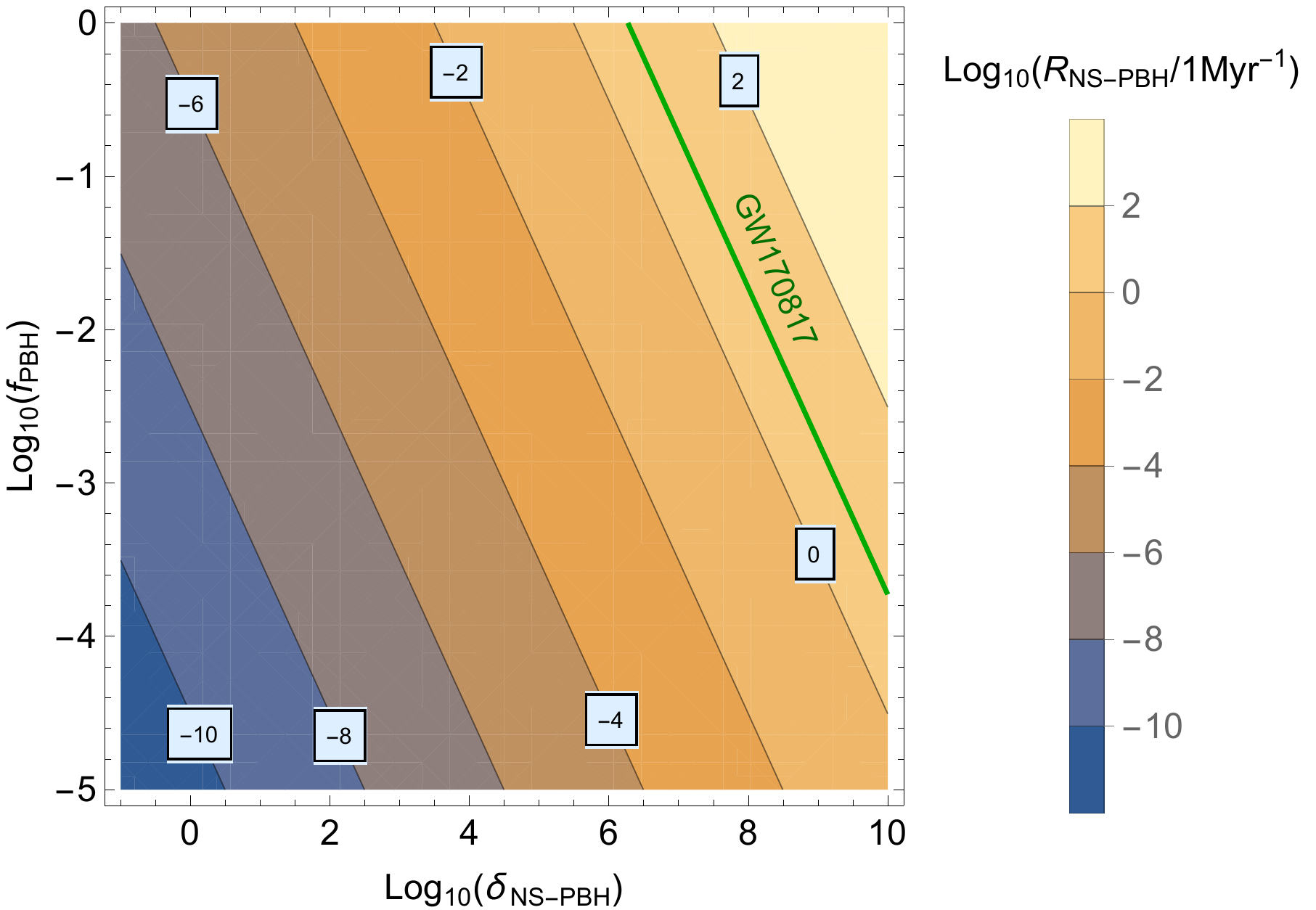}
    \caption{Contours of constant NS-PBH merger rate $R_{\rm NS-PBH}$ in the plane defined by the NS-PBH density contrast $\delta_{\rm NS-PBH}$ versus the fraction of dark matter $f_{\rm PBH}$ that is in the form of PBH. The green line indicates the rate needed for GW170817 to be a potential NS-PBH merger.}
    \label{fig:NSPBH}
\end{figure}

Assuming that $\rho_1=\rho_{\rm NS}$ from Eq. (\ref{eq:nsdist}), $\rho_2=\rho_{\rm PBH}$ as described in \ref{sec:DM_dist}, $m_1=1\ M_\odot$ and $m_2=1\ M_\odot$, we find that, for a galaxy with the properties described above, the rate for mergers of primordial black holes and neutron stars (PBH-NS) is 
\begin{equation}\label{eq:pbhns}
R_{\rm NS-PBH}\simeq3.2\times 10^{-6}f_{\rm PBH}\delta_{\rm NS-PBH}\: {\rm Myr}^{-1},
\end{equation}
where $f_{\rm PBH}$ is the fraction of the dark matter in PBH (considered to be constant throughout the universe), and where $\delta_{\rm NS-PBH}$ is a density contrast factor accounting for the possibility of finding more neutron stars in a dark matter-rich environment. Notice that varying the merging bodies' masses within the range for GW170817 we find a variation in the rate smaller than 5\% and thus negligible comparing to other sources of uncertainty. In Fig. \ref{fig:NSPBH} we show our results for the PBH-NS rate for different values of $f_{\rm PBH}$ and $\delta_{\rm NS-PBH}$.

\begin{figure}
    \centering
    \includegraphics[width=9.0 cm]{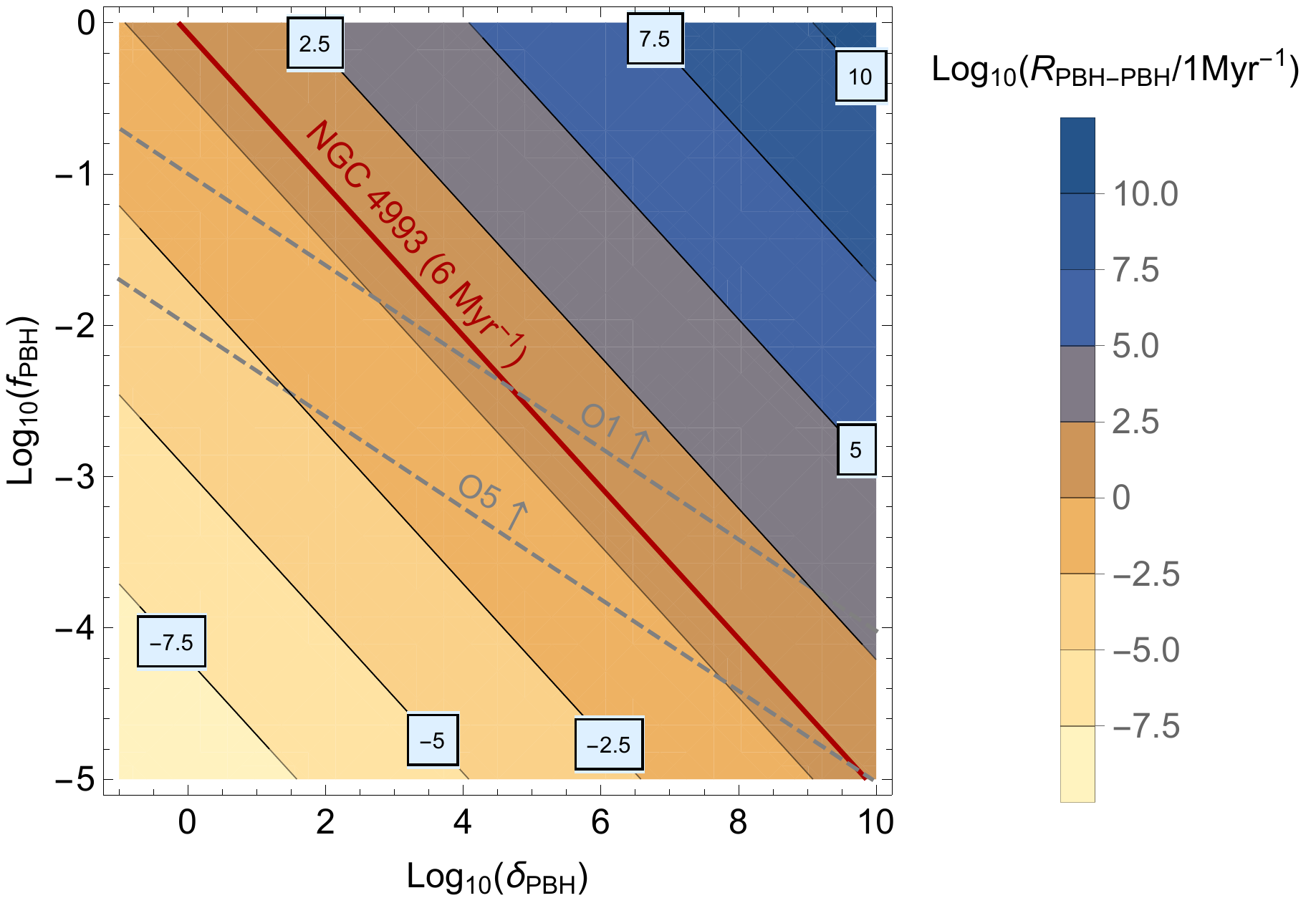}
    \caption{Contours of constant PBH-PBH merger rate $R_{\rm PBH-PBH}$ in the plane defined by the PBH density contrast $\delta_{\rm PBH}$ versus the fraction of dark matter $f_{\rm PBH}$ that is in the form of PBH. The red line indicates a $6~ \rm Myr^{-1}$ rate in NGC 4993, comparable to the rate estimation for normal field BNS \cite{palmese}. The dashed gray lines are potential constraints (the arrows point to the constrained regimes) on these parameters set by non-observation of PBH-PBH mergers \cite{Raidal:2017mfl}, under a rather strong assumption, that the density contrast is the same in the early universe, in the late universe, and in NGC 4993, as discussed in Section \ref{sec:dynamical}. The line labeled ``O1'' is derived in \cite{Raidal:2017mfl} using the LIGO/Virgo data from the first observing run, while the other constraint is a projection for the fifth observing run.}
    \label{fig:PBH}
\end{figure}

For comparison, we find that the PBH-PBH merger rate is
\begin{equation}
   R_{\rm PBH}\simeq8.3\ f_{\rm PBH}^2\delta_{\rm PBH}\:{\rm Myr}^{-1}, 
\end{equation} 
while the
BNS merger rate for this dynamical formation scenario is
\begin{equation}\label{eq:nsns}
    R_{\rm BNS, dyn}\simeq4.4\times 10^{-10}\delta_{\rm NS}\: {\rm Myr}^{-1},
\end{equation}
 where $\delta$ factors represent the enhancement from density contrast \cite{Raidal:2017mfl}. The results of the PBH-PBH merger-rate calculation are shown Fig. \ref{fig:PBH} for different values of $f_{\rm PBH}$ and $\delta_{\rm PBH}$.

Notice that varying the NS population parameter $R_0$ affects the rates in Eq.~(\ref{eq:pbhns}) and (\ref{eq:nsns}) very marginally, to within a factor 3 for variations of $R_0$ within a factor 2. On the other hand, the rate in Eq.~(\ref{eq:pbhns}) is directly proportional to the total number of NS, $N_{\rm NS}$, while the rate in Eq. (\ref{eq:nsns}) depends quadratically on it.

For what concerns the DM portion of the rate calculation, we find that the most critical parameter is the DM halo concentration. However, changes to the rate are minimal. Setting the concentration to its central value plus 1$\sigma$ only amplifies the NS-PBH merger rate by around 50\%, while leaving the PBH-PBH merger rate almost unchanged. A concentration set at $c_{200}=10$ gives, for instance,
\begin{equation}
    R^{c_{200}=10}_{\rm PBH-NS}=4.2\times 10^{-6}\ f_{\rm PBH}\ \delta_{\rm NS-PBH}\: {\rm Myr}^{-1}.
\end{equation}

In order for the NS-PBH scenario to be plausible, we require the NS-PBH rate to be at least as high as the field binary formation rate in this galaxy, around 6 Myr$^{-1}$. In addition, assuming that there has not been a detection of a PBH-PBH merger event (though GW190425 is a potential candidate), it is a fair assumption for internal consistency to require $R_{\rm PBH}\sim R_{\rm NS-PBH}$ for the PBH-PBH rate not to be much higher than the NS-PBH rate. Additionally, 
Using our benchmark choices for the various relevant parameters, in order to obtain the condition $R_{\rm NS-PBH}\sim R_{\rm PBH}\sim 6\ {\rm Myr}^{-1}$, we need the following: \begin{equation}
    \quad \delta_{\rm NS-PBH}\sim \frac{2\times 10^6}{f_{\rm PBH}}\:;\;\;\;\delta_{\rm PBH}\sim\frac{1}{f_{\rm PBH}^2}.
\end{equation} For instance, for $f_{\rm PBH}\sim 10^{-1}$, we need $\delta_{\rm NS-PBH}\sim 2\times 10^7$ and $\delta_{\rm PBH}\sim 10^2$. For $f_{\rm PBH}\sim 10^{-2}$, we need $\delta_{\rm NS-PBH}\sim 2\times 10^8$ and $\delta_{\rm PBH}\sim 10^4$. The values of $\delta_{\rm PBH}$ considered in the literature vary wildly and can be well in excess of $10^4$ (see e.g. \cite{Clesse_2017,Clesse:2016ajp}, which consider values of $\delta_{\rm PBH}\sim10^{10}$). The large values needed in our scenario for $\delta_{\rm NS-PBH}\sim 10^7$ to $10^9$ 
are troubling at face value: they would reflect a significant bias for neutron stars to be present in large dark matter overdensities. However, this is not entirely implausible for the following reasons: large dark matter density contrasts on the order of $10^9$ to $10^{10}$ are observed, and are indeed typical in globular clusters and ultra-faint dwarf galaxies observed by Keck/DEIMOS \cite{Martin:2007ic,Simon:2007dq} and the Dark Energy Survey \cite{Drlica-Wagner:2015ufc}. 
Additionally, globular clusters are known to host large densities of NS, due to to stellar encounters that should occur frequently in environments of high stellar density \cite{1975ApJ...199L.143C}; observations of low-mass X-ray binaries and millisecond pulsars confirm this. We conclude that it is not implausible that concurrent overdensities in dark matter and NS density produce $\delta_{\rm NS-PBH}\sim 10^7 - 10^9$. Additionally, recent merger history could also enhance the expected value for $\delta_{\rm NS-PBH}$.

Our calculations involving NS natal kick and velocity dispersion show that the rate can be enhanced by an order of magnitude (a factor of $\sim$ 7, to be exact), but would not change the conclusions of this section.

\begin{figure}
    \centering
    \includegraphics[width=8.5 cm]{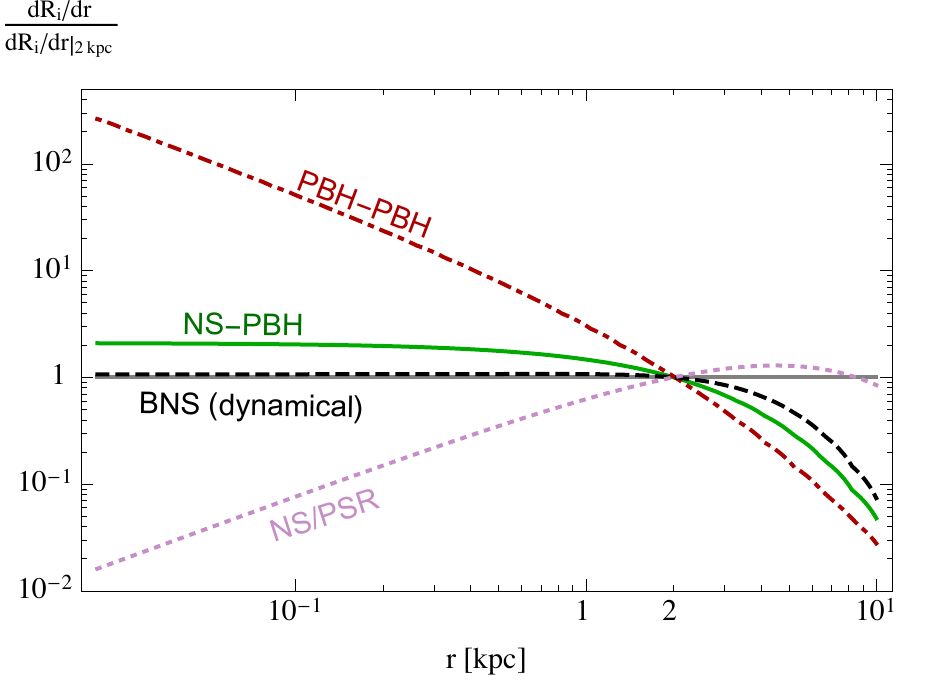}
    \caption{Projected radial distribution of the dynamical binary formation rates for both PBH-PBH (dot--dashed red) and NS-PBH (solid green) mergers, with the rate normalized to one at 2 kpc, the distance of the GW170817 counterpart to the center of NGC 4993. For comparison, we also show the BNS rate distribution from the dynamical formation as computed in this work (dashed black), and the NS distribution assumed in our analysis (dotted purple).
    } 
    \label{fig:radial}
\end{figure}

An additional source of information is the specific location of the kilonova AT2017gfo associated with GW170817, offset by around 2 kpc from the galaxy center. In Fig. \ref{fig:radial} we show the projected radial distribution of the rates presented in this section (i.e. the ratio of integrals along a line of sight of the merger rate). Note that the expected distribution of NS-PBH mergers has a distinctive shape that in the case of NGC 4993 is roughly flat out to $\sim 2$ kpc, which corresponds to the distance of the kilonova AT2017gfo from the center of the galaxy. It is therefore likely that, if GW170817 was a NS-PBH merger, it happened within this projected distance from the center. In the future, it will be interesting to compare these distributions to Gamma--Ray Bursts (GRBs) and GW counterparts locations to confirm or rule out this binary formation scenario \cite{precite}.

In this section, we use an analytical estimation to determine the merger rates of dynamically formed NS-PBH and PBH-PBH mergers through gravitational interactions. A sophisticated merger rate determination requires considerations of the detailed N-body dynamics, stellar evolution, environmental effects (as done for the compact mergers, e.g. \cite{OLeary:2008myb,thompson2010accelerating,Lee_2010,East:2012ww,Rodriguez:2016avt}).
We leave this for future works \cite{precite}.

\section{NS Implosion by Micro-sized PBH}
\label{sec:implosion}

\begin{figure}
\centering
\includegraphics[width= 8.5 cm]{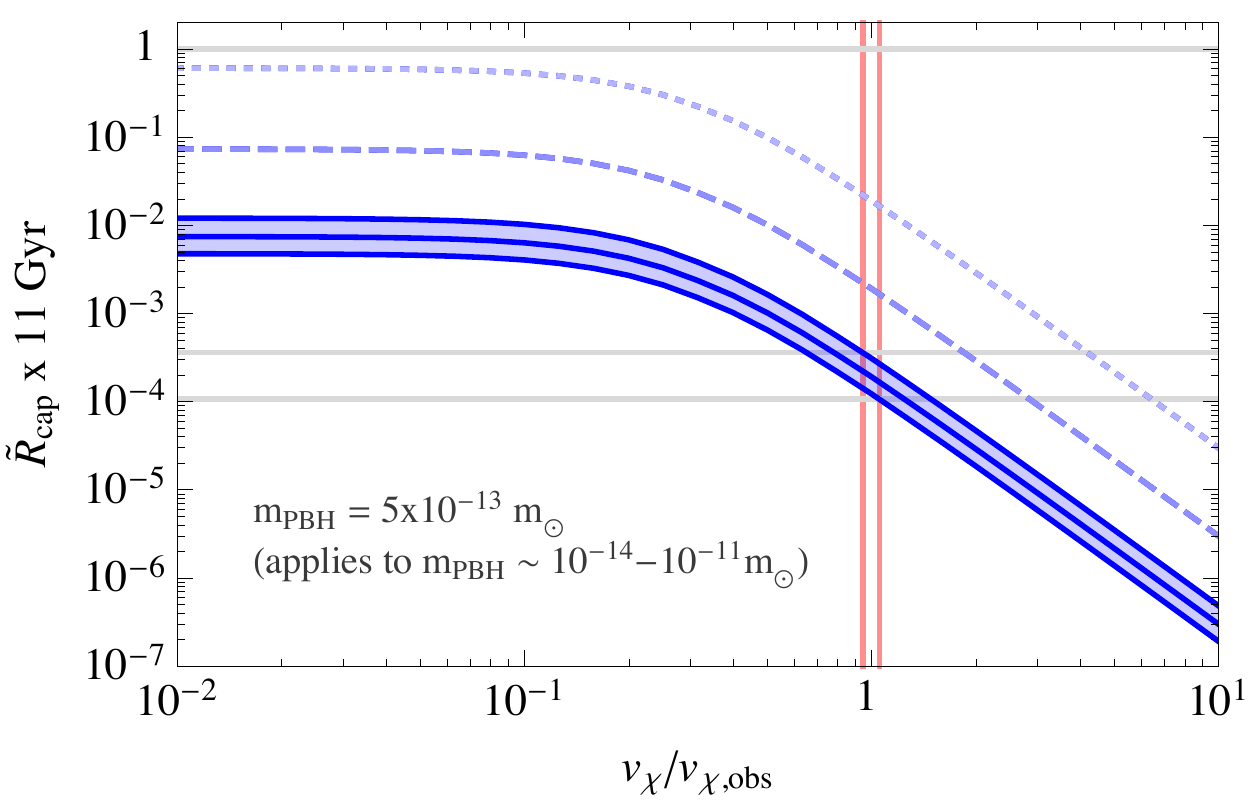}
\caption{
The capture rate of PBH by either one of the NS, times 11 Gyr, in terms of the ratio of velocity dispersion $v_\chi/v_{\chi, \rm obs}$ (normalized to its observed central value from NGC 4993 $v_{\chi, \rm obs}$).
The blue band corresponds to the values for $\rho_{\rm pbh} = \rho_{\rm DM, obs}$ within the error discussed in \cite{Ebrova:2018gtz}. The blue dashed curve and the blue dotted curves assumes an enhanced DM abundance $10 \times \rho_{\rm DM, obs}$ and $100 \times \rho_{\rm DM, obs}$ (simply in preparation for more extreme environments for future mergers), respectively, using the central value of $\rho_{\rm DM, obs}$.
We also use the two vertical lines to label the x-axis value of $(v_{\chi,{\rm obs}} \pm 9.1~ {\rm km~s^{-1}})/v_{\chi, {\rm obs}}$.
The lower gray curves indicate the expected number of events, $1.9^{+1.7}_{-0.9} \times 10^{-4}$, assuming the best-known parameters for the GW170817 location. 
Though we choose $\rm m_{pbh}= 5\times 10^{-13} m_\odot$ to make the plot, the result applies to $\rm m_{pbh}\sim 10^{-14} - 10^{-11} m_\odot$ (for details, see Sec. \ref{sec:implosion}.)
}
\label{fig:implosion}
\end{figure}

In this section, we study a different scenario, considering how likely it is that a NS is imploded into a light BH, 2 kpc away from the center of NGC 4993, by capturing a micro-sized PBH, and resulting in GW170817 as a NSBH merger. Here, we label the tiny PBH as "pbh" in equations and plots to distinguish it from the discussions in previous sections.

An interesting mass range to consider for the PBH is $10^{-15}$ to $10^{-8}$ solar mass, as discussed in \cite{roncadelli2009primordial,Carr:2009jm, Capela:2013yf,Pani:2014rca,Carr:2020gox}. The radius of the PBH is about $r \sim 10^{-8} \rm cm\;(m_{\rm pbh}/10^{-13} m_\odot)$.
They have been considered to be captured by or to destroy white dwarfs (WD) \cite{Capela:2012jz} and neutron stars \cite{Capela:2013yf}.
The potential constraints set by these considerations are sensitive to the assumptions of the properties of each astrophysical systems, which have large observational uncertainties 
\cite{Bradford_2011,Ibata_2012,Capela:2014qea,Defillon:2014wla,Genolini:2020ejw}.
Given that the strong Subaru Hyper Suprime-Cam (HSC) constraints \cite{Niikura:2017zjd}
kick in at $\gtrsim$ $10^{-11} m_\odot$
and the NS consumption rate drops rapidly below $10^{-14} m_\odot$ \cite{Capela:2013yf}, we further restrict our attention to the mass range of $\sim 10^{-14} - 10^{-11} m_\odot.$
Note that previous studies of the NS constraint in this mass range are highly sensitive to the assumed DM abundance in globular clusters, and the constraint was drawn assuming high DM abundance (up to $10^4\;\rm GeV\:cm^{-3}$) \cite{Capela:2013yf}. A potential constraint on PBH triggering supernovae for WD \cite{Graham:2015apa} is also being heavily debated \cite{Montero-Camacho:2019jte}. Further investigations of these constraints are beyond the scope of this paper.

Here, we assume PBHs account for all of the DM abundance in NGC 4993, but the analysis can be easily adjusted for different assumptions of PBH abundance. 
The scenario where a NS is imploded by PBH through capturing is considered in \cite{Capela:2013yf}. 
PBHs with mass density $\rho_{\rm pbh}$ can be captured by NS at an estimated rate \cite{Capela:2013yf}, 
\begin{align}
R_0=\sqrt{6\pi}\frac{\rho_{\rm pbh}}{m_{\rm pbh}}\left(\frac{2 G M R}{v_{\rm x}}\right) \frac{1-{\rm Exp}\left[-\frac{3E_{\rm loss}}{m_{\rm pbh}v_{\rm x}^2}\right]}{1-\frac{2GM}{R}} \, ,
\label{eq:pbhcap}
\end{align}
where $M$ is the NS mass (which will be specified later), and $R$ is the NS radius, assumed to be 12 km (note that the conclusion is insensitive to this assumption). $v_x$ is the velocity dispersion of the PBH.
$E_{\rm loss} \simeq \frac{4 G^2 m^2_{\rm pbh}M}{R^2} \left\langle  \frac{\ln\Lambda}{2 G M / R} \right\rangle$ is the average energy loss of a PBH in the NS assuming a constant flux. $ \left\langle \frac{\ln\Lambda}{2 G M / R}\right\rangle \sim 14.7$ for a typical neutron star \cite{Belvedere:2012uc} (the Coulomb logarithm $\ln(\Lambda)$ is discussed in \cite{1987gady.book.....B,Capela:2013yf}).

We define $R_{\rm capture}$ as the modified rate based on $R_0$, folding in both the velocity distributions of PBH and NS (as discussed in \cite{Cordes:1997my,Fuller:2017uyd}),
\begin{align}
R_{\rm capture} = & \int d^3v_{\rm ns} f^{3D}_{\rm ns} (\vec{v}_{\rm ns}, \overline{v}_{\rm ns}) \int d^3v_{\rm pbh} f^{3D}_{\rm pbh} (\vec{v}_{\rm pbh}, \overline{v}_{\rm pbh}) 
\nonumber
\\
&\times
R_0 (|\vec{v}_{\rm ns} - \vec{v}_{\rm pbh}|)~.
\end{align}
Here, $\rm ns$ and $\rm pbh$ denote NS and PBH, respectively. 
$f^{3D} (\vec{v}, \overline{v})$ is the 3-D Maxwellian velocity distribution and $\overline{v}$ is the mean of $|\vec{v}|$.
$|\vec{v}_{\rm ns} - \vec{v}_{\rm pbh}| \equiv (v_{\rm ns}^2 + v_{\rm pbh}^2 - 2 v_{\rm ns} v_{\rm pbh} \cos \theta)^{1/2}$, and $\theta$ is the angle between velocities. For the weighting integrand, we use the 1-D Maxwellian distribution
\begin{equation}
f^{1D}(|\vec{v}|,\overline{v}) = \left(\dfrac{3}{2 \pi \overline{|\vec{v}|}^2}\right)^{3/2} e^{- 3 |\vec{v}|^2/(2\overline{v}^2)}
\end{equation} to approximate the dependence on the radial component of the 3D velocities. More details can be found in the Appendix of \cite{Fuller:2017uyd}.
Finally, we define $\widetilde{R}_{\rm cap}$ as the rate of capturing a PBH by one of the NS's in a binary, but not both, and use individual NS masses as inputs in the calculation.

Our benchmark PBH mass is $\sim 5\times 10^{-13} m_\odot$, but the capture rate is almost flat in the range of $\sim$ $10^{-14}$ to $10^{-11} m_\odot$ assuming a fixed DM mass density and velocity dispersion.
The expected number of events of PBH being captured by a NS in 11 Gyr is roughly $1.9^{+1.7}_{-0.9} \times 10^{-4}$. 
This means the capture and implosion scenario is extremely unlikely to produce a NS-BH merger as the GW170817 event.
Our result is consistent with \cite{Bramante:2017ulk}, which considers PBH-NS implosion event rates in the Milky Way; the implosion rate is extremely small given a realistic PBH-DM mass density and NS number density. However, given a higher density or lower DM velocity than what is assumed here, the implosion could still have a significant rate and can create the exotic low mass BH-NS merger through this implosion scenario.

In Fig. \ref{fig:implosion}, we plot the expected rate for one NS (but not both) in a binary to capture a PBH, $\widetilde{R}_{\rm cap}$, times 11 Gyr, as a function of the velocity dispersion $v_\chi$ normalized to its observed central value $v_{\chi, {\rm obs}}\equiv 160.0 ~ {\rm km~s^{-1}}$ in NGC 4993.
Note that the PBH is to assumed to have mass $ 5\times 10^{-13} m_\odot$, and the NS is considered to be located 2 kpc away from the center of NGC.
The blue band is the capture rate times 11 Gyr (roughly the age of NGC 4993) depending on $v_\chi/v_{\chi, \rm obs}$, considering $\rho_{\rm pbh} = \rho_{\rm DM, obs}$, and $\rho_{\rm DM, obs}$ is derived from $M_{\rm halo} = 194^{+120}_{-70} \times 10^{10} M_\odot$ \cite{Ebrova:2018gtz}. See more discussions in Sec. \ref{Sec:NGC_4993}.
We also draw two light red curves indicating the 1$\sigma$ uncertainty on $v_{\chi, {\rm obs}}$.
The blue-dashed/dotted curves are obtained assuming 10 and 100 times larger PBH-DM abundance than what is derived in \cite{Ebrova:2018gtz}, respectively. 
We also add gray horizontal lines to help visualize where the rate is unity, and where the expected number of events is $1.9^{+1.7}_{-0.9} \times 10^{-4}$, given the uncertainties in the DM properties in NGC 4993.

The curves of varying DM density and velocity dispersion help motivate future implosion studies, if there is a GW170817-like event happening in a galaxy with higher DM density and lower DM velocity dispersion comparing to NGC 4993. As mentioned, the observed events GW190425 \cite{190425} and GW190814 \cite{Abbott:2020khf} (involving heavy BH) could have the potential to be such candidates, but their analysis is challenging given the lack of optical signatures and information on the host galaxy.

The NS implosion can also be induced by asymmetric dark matter (ADM) accumulating in the NS \cite{Goldman:1989nd,deLavallaz:2010wp,Kouvaris:2010jy,McDermott:2011jp,Bell:2013xk,Petraki:2013wwa,Zurek:2013wia,Bramante:2017ulk,Kouvaris:2018wnh}. 
We leave the consideration of ADM induced implosion for GW170817 and NGC 4993 to future works, given the complication in calculating capture rates depending on the ADM masses and interactions. 
PBH-induced implosion has the advantage of having the minimal assumption that the object is simply a microsized BH and is not subject to the complication of the variety of the ADM interactions assumed in the literature.

\section{Conclusions and Prospects for GW190425 $\&$ GW190814}
\label{sec:prospects}

In this work, we explored the possibility that GW170817 is a compact-object merger that involves a PBH. Starting from the observed properties of the host galaxy, NGC 4993, we estimated the expected rates of NS--PBH, PBH--PBH and BNS mergers dynamically formed through two-body gravitational interactions in NGC 4993.
We find that the rate for NS--PBH and PBH-PBH mergers are generally significantly higher than that for dynamically formed BNS if one assumes density contrast factors of the same order, and if PBHs are a significant fraction of the DM.
The NS-PBH scenario could provide a significant contribution to the standard binary formation rate, $\sim 6 ~\rm Myr^{-1}$, inferred from NGC 4993, with a large boost factor $\delta_{\rm NS-PBH}$.
Namely, we need $\delta_{\rm NS-PBH}\sim {2\times 10^6}/{f_{\rm PBH}}$ in order for this scenario to be competitive with the isolated binary scenario predicted by stellar evolution in NGC 4993.

We also provided the expected radial distribution of NS-PBH/PBH-PBH/dynamical BNS mergers. It will be interesting to compare the distribution of observed BNS/NS-BH mergers with our model to discern between different formation channels, similar to the study conducted in \cite{Bramante:2017ulk} to distinguish ADM imploded BNS distribution from the usual BNS distribution, in future works \cite{precite}.
The dynamical formation scenario involving PBH suggested here provides a new formation channel to account for the volumetric rate reported by LIGO/Virgo, as discussed more specifically below. 

Observationally, LIGO/Virgo have reported various events of interest for our exotic scenarios.
The event GW190425 is peculiar because the total mass ($\sim 3.4 ~M_\odot$) of the system is significantly larger \citep{190425} than other known BNS systems. Moreover, no compelling EM counterparts have been found. While finding a counterpart at the distance of this event ($159^{+69}_{-71} $ Mpc) with a poor spatial localization ($\sim 8,000$ sq deg) can prove extremely challenging, a consequence is that the possibility that this is a light BBH or a NSBH merger cannot be excluded.

Due to the lack of understanding of its host galaxy, one cannot conduct as detailed an analysis as what we performed in this paper for GW190425. However, the assumptions made in this work about the DM and NS distribution of NGC 4993 are quite general, and we can potentially extend some of our conclusions to this new event. In fact, our predicted NS-PBH or PBH-PBH rates mostly depend on the DM concentration parameter, and also on the total number of NS in the galaxy for NS-PBH. This implies that if GW190425 includes a PBH, it is more likely to have happened in a galaxy with large stellar mass and/or massive and concentrated DM halo.  LIGO/Virgo find that the rate of GW190425--like events is $460^{+1050}_{-390}~{\rm Mpc}^{-3}$ yr$^{-1}$ \cite{190425} and one can use our NS-PBH and PBH-PBH merger rates to account for that, summing over the potential contributing galaxies in an appropriate volume. 

More recently, a peculiar compact-object binary merger, GW190814 \cite{Abbott:2020khf}, was detected in gravitational waves by LIGO/Virgo. This event is also of great interest for this work because it originated from a binary, most likely a BBH, where the secondary has a mass of $\sim 2.6 M_\odot$. If the secondary is a BH, it would be the lightest BH ever observed in a binary, and it is challenging for current BBH formation theories to produce a merger with the properties of GW190814. One could again explore the possibility of GW190814 containing a PBH as its secondary \cite{Vattis:2020iuz}, and conduct an analysis similar to this work by making assumptions about its host galaxy \cite{precite}.

In the second part of this work, we explore the possibility that GW170817 is a NSBH merger, where the BH was originally a NS imploded by a microscopic PBH. We find that this scenario is very unlikely for the location of GW170817 in the environment of NGC 4993, but such a scenario could happen in galaxies with large DM density and small DM velocity dispersion.

Our study provides motivation for further studies to distinguish between BNS, NSBH, and BBH with solar-mass component objects from their GW signal \cite{Yang:2017gfb,Chatziioannou:2018vzf,Chen:2019aiw,Fasano:2020eum,Chen:2020fzm}. It also motivates a detailed analysis taking into account a range of galaxy types with different DM and NS environments, especially for a comparison with the volumetric rates of BBH, BNS, and NSBH as measured by LIGO/Virgo \cite{precite}.

\newpage

\acknowledgments

We thank Andrew Levan, Jens Hjorth, and Joe Lyman for sharing the MUSE velocity dispersion maps. We also thank Dan Hooper, Mariangela Lisanti and Volodymyr Takhistov for useful discussions.
This document was prepared by Y-DT and AP using the resources of the Fermi National Accelerator Laboratory (Fermilab), a U.S. Department of Energy, Office of Science, HEP User Facility. Fermilab is managed by Fermi Research Alliance, LLC (FRA), acting under Contract No. DE-AC02-07CH11359. 
SP is partly supported by the U.S. Department of Energy grant number de-sc0010107.
This work was partially conceived at the Aspen Center for Physics, which is supported by National Science Foundation grant PHY-1607611.

\bibliographystyle{hunsrt.bst}
\bibliography{biblio}

\end{document}